\documentclass[twocolumn,amsmath,showpacs,nofootinbib]{revtex4-1}
\usepackage{graphicx}
\usepackage{dcolumn}
\usepackage{bm}
\usepackage{color} 
\usepackage{slashed}
\begin{document}
\newcommand{\hs}{\hspace*{0.5cm}}
\newcommand{\vs}{\vspace*{0.5cm}}
\newcommand{\be}{\begin{equation}}
\newcommand{\ee}{\end{equation}}
\newcommand{\bea}{\begin{eqnarray}}
\newcommand{\eea}{\end{eqnarray}}
\newcommand{\ben}{\begin{enumerate}}
\newcommand{\een}{\end{enumerate}}
\newcommand{\bde}{\begin{widetext}}
\newcommand{\ede}{\end{widetext}}
\newcommand{\nn}{\nonumber}
\newcommand{\crn}{\nonumber \\}
\newcommand{\Tr}{\mathrm{Tr}}
\newcommand{\non}{\nonumber}
\newcommand{\noi}{\noindent}
\newcommand{\al}{\alpha}
\newcommand{\la}{\lambda}
\newcommand{\bet}{\beta}
\newcommand{\ga}{\gamma}
\newcommand{\va}{\varphi}
\newcommand{\om}{\omega}
\newcommand{\pa}{\partial}
\newcommand{\+}{\dagger}
\newcommand{\fr}{\frac}
\newcommand{\bc}{\begin{center}}
\newcommand{\ec}{\end{center}}
\newcommand{\Ga}{\Gamma}
\newcommand{\de}{\delta}
\newcommand{\De}{\Delta}
\newcommand{\ep}{\epsilon}
\newcommand{\varep}{\varepsilon}
\newcommand{\ka}{\kappa}
\newcommand{\La}{\Lambda}
\newcommand{\si}{\sigma}
\newcommand{\Si}{\Sigma}
\newcommand{\ta}{\tau}
\newcommand{\up}{\upsilon}
\newcommand{\Up}{\Upsilon}
\newcommand{\ze}{\zeta}
\newcommand{\ps}{\psi}
\newcommand{\Ps}{\Psi}
\newcommand{\ph}{\phi}
\newcommand{\vph}{\varphi}
\newcommand{\Ph}{\Phi}
\newcommand{\Om}{\Omega}

\title{Scotogenic gauge mechanism for neutrino mass and dark matter} 

\author{Phung Van Dong} 
\email{dong.phungvan@phenikaa-uni.edu.vn}
\affiliation{Phenikaa Institute for Advanced Study, Phenikaa University, Yen Nghia, Ha Dong, Hanoi, Vietnam}
\author{Nguyen Huy Thao}
\email{nguyenhuythao@hpu2.edu.vn (corresponding author)}
\affiliation{Department of Physics, Hanoi Pedagogical University 2, Phuc Yen, Vinh Phuc, Vietnam} 

\date{\today}

\begin{abstract}
Scotogenic is a scheme for neutrino mass generation through the one-loop contribution of an inert scalar doublet and three sterile neutrinos. This work argues that such inert scalar doublet is a Goldstone boson mode associated with a gauge symmetry breaking. Hence, the resultant scotogenic gauge mechanism is very predictive, generating neutrino mass as contributed by a new gauge boson doublet that eats such Goldstone bosons. The dark matter stability is manifestly ensured by a matter parity as residual gauge symmetry for which a vector dark matter candidate is hinted.                   
\end{abstract} 

\maketitle

\section{Motivation} 

Neutrino mass and dark matter are among the crucial questions in science, which cannot be solved within the conventional framework of the standard model \cite{ParticleDataGroup:2022pth}. 

The scotogenic setup is perhaps the most compelling proposal in explaining both of the issues \cite{Ma:2006km,Tao:1996vb}. Indeed, this model introduces an inert scalar doublet and three sterile neutrinos that couple to usual lepton doublets, producing relevant neutrino masses entirely through one-loop corrections of both types of these new fields. Intriguingly, the lightest inert scalar or sterile neutrino field is availably to be a dark matter candidate. However, an exact $Z_2$ symmetry that assigns all the new fields to be odd, while putting the usual fields to be even, is necessary for the model properly working as well as stabilizing the dark matter candidate. The nature origin of the $Z_2$ symmetry has been left as an open question. This matter has been extensively studied in \cite{Ma:2013yga,Ma:2013yga,Ma:2016nnn,Kang:2019sab}.

We would like to suggest here that the fundamental symmetry principle for $Z_2$ can be understood by a gauge completion for the scotogenic mechanism. This gauge completion is commonly hinted from numerous potential extensions of the standard model, such as 3-3-1 model and trinification, which actually lead to such a residual $Z_2$ symmetry \cite{Dong:2013wca,Dong:2017zxo}. Interestingly, the inert scalar doublet that is odd under $Z_2$ becomes the Goldstone boson mode of a new gauge boson doublet (see \cite{VanDong:2021xws} for a non-gauge version of it). This new vector doublet taking part in the extended gauge field is odd under $Z_2$ too. Consequently, the gauge vector doublet governs neutrino mass generation as well as providing a dark matter candidate, instead of the inert scalar candidate in the normal sense. That said, the model governed by the gauge principle is very predictive, opposite to the original approach. 

The rest of this work is as follows. In Sec. \ref{intro}, we give a review of the scotogenic mechanism. In Sec. \ref{sgm}, we propose the scotogenic gauge mechanism. In Sec. \ref{sgc}, we discuss a gauge completion possibility, giving the $Z_2$. In Sec. \ref{dmlfv}, we make remarks about dark matter observables and charged-lepton flavor violation processes. We conclude this work in Sec. \ref{concl}.  

\section{\label{intro} Brief description of the scotogenic mechanism}            

The standard model arranges left-handed fermions in doublets, $l_{aL}=(\nu_{aL},e_{aL})$ and $q_{aL}=(u_{aL},d_{aL})$, while it puts right-handed fermions in singlets, $e_{aR}$, $u_{aR}$, and $d_{aR}$, where $a=1,2,3$ is a family index. A Higgs doublet $\varphi =[G^+_W,(v+H+iG_Z)/\sqrt{2}]$ for gauge symmetry breaking and mass generation is required, while no right-handed neutrino partners are presented. 

The scotogenic model~\cite{Ma:2006km,Tao:1996vb} introduces three right-handed neutrino singlets $\nu_{i R}$ ($i=1,2,3$) and an inert Higgs doublet $\eta=(\eta^0,\eta^-)$, which all are odd under a $Z_2$ symmetry, whereas the standard model fields are even under this group. Consequently, the Yukawa Lagrangian includes 
\be \mathcal{L}\supset h_{ai} \bar{l}_{aL} \eta \nu_{iR}-\fr 1 2 M_i \nu_{i R}\nu_{i R}+H.c.,\ee while the scalar potential contains
\bea V &\supset& \mu^2_\eta \eta^\dagger \eta +\fr 1 2  \la_\eta (\eta^\dagger \eta)^2\crn
&&+ \la_1 (\eta^\dagger \eta) (\varphi^\dagger \varphi)+ \la_2(\varphi^\dagger \eta)(\eta^\dagger \varphi)\crn
&&+\fr 1 2 \la_3 [(\eta\varphi)^2+H.c.]. \eea Above, we assume $\nu_{iR}$ to be a physical Majorana field by itself with mass $M_i$ for each $i=1,2,3$. It is clear that $\eta=[(S+i A)/\sqrt{2},\eta^-]$ does not have a VEV for $\mu^2_\eta>0$ and is separated in mass, such as
\bea   m^2_{\eta^\mp} &=& \mu^2_\eta + \la_1 v^2/2,\\ 
m^2_{S}&=& \mu^2_\eta + (\la_1+\la_2 +\la_3)v^2/2,\\
m^2_{A}&=& \mu^2_\eta + (\la_1+\la_2 - \la_3)v^2/2.\eea

\begin{figure}[h]
\bc
\includegraphics[scale=1]{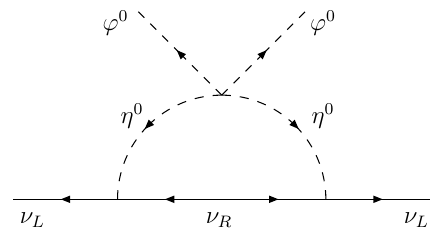}
\caption[]{\label{fig0} Scotogenic neutrino mass generation.}
\ec
\end{figure}

The neutrino mass generation diagram is given in Fig.~\ref{fig0}. Hence, we obtain radiative neutrino masses,
\bea [m_{\nu}]_{ab} &=& \fr{h_{ai}h_{bi} M_i}{32\pi^2}\left[\fr{m^2_S}{m^2_S-M^2_i}\ln \fr{m^2_S}{M^2_i}\right.\crn
&&\left.-\fr{m^2_A}{m^2_A-M^2_i}\ln \fr{m^2_A}{M^2_i}\right],\eea where $i$ is summed over $i=1,2,3$. Because the $S$-$A$ mass splitting is $m^2_S-m^2_A=\la_3 v^2\ll m^2_0\equiv (m^2_S+m^2_A)/2\sim M^2_i$, it leads to \be [m_\nu]_{ab}\sim \fr{\la_3 v^2}{32\pi^2}\fr{h_{ai}h_{bi}}{M_i}.\ee Hence, the seesaw scale $M_i$ is suppressed by a scalar coupling and loop factor, i.e. $\la_3/32\pi^2$, opposite to canonical seesaw. This prediction requires $M_i$ not necessarily to be large. Indeed, it matches the observation, $m_\nu \sim 0.1$ eV, given that $\la_3\sim h^2 \sim 10^{-4}$ and $M_i$ at TeV. 

An intriguing result is that the lightest of $S,A,\nu_{iR}$, which contribute to neutrino masses, is stabilized by $Z_2$ conservation responsible for dark matter. This dark matter candidate, which is either a $S,A$ scalar or a $\nu_{iR}$ fermion, has been extensively studied in~\cite{Escribano:2020iqq,VanDong:2023thb,VanDong:2023xbd}. The implication of the scotogenic mechanism is that neutrino mass is induced by dark matter.   

Above, the condition $\mu^2_\eta>0$ is important. First, it ensures $\langle \eta\rangle =0$, i.e. $Z_2$ is not broken by the vacuum as desirable. Second, $\eta$ is a physical degree of freedom, i.e. the scotogenic scheme is completed on its own. However, what happens if $\mu^2_\eta<0$? Given the only standard model gauge symmetry, $\eta$ develops a VEV that breaks $Z_2$; additionally, a part of $\eta$ becomes Goldstone bosons of $W,Z$ as mixed with the Higgs $\varphi$, analogous to the two Higgs doublet model. The conventional scotogenic mechanism is ruined, since the dark matter fast decays, while the neutrinos get an unavoidably large tree-level mass. The new observation of this work is that if a gauge completion for the scotogenic scheme with $\mu^2_\eta<0$ exists, $\eta$ becomes the Goldstone mode of a new vector doublet, which has vanished VEV and mass $\mu^2_\eta+\la_\eta w^2/2 =0,$\footnote{Similar to that condition for $G^\pm_W$ in the standard model.} protected by the new gauge symmetry (i.e. Goldstone theorem), where $w$ is the gauge completion scale. The $Z_2$ symmetry is restored, originating from the new gauge symmetry. Neutrino mass is contributed by the vector doublet instead of its Goldstone $\eta$. In what follows, this hypothesis is explicitly proved. The statement of gauge completion for scotogenic mechanism is always understood as adjusting relevant potential mass term to be negative.          

\section{\label{sgm} Proposal of the scotogenic gauge mechanism} 

Assuming a simple gauge group $\mathcal{S}$, which is spontaneously broken by a heavy Higgs field $\chi^0$ down to the $SU(2)_L$ group, supplies a massive gauge boson doublet $V_\mu$ by eating a relevant Goldstone boson mode $\eta$, such as     
\bea V_\mu &=& \left[\begin{array}{c}
V^0_\mu \\
V^-_\mu \end{array}\right]=\left[\begin{array}{c}
\fr{1}{\sqrt{2}}(V_{1\mu}+iV_{2\mu}) \\
V^-_\mu \end{array}\right],\\ 
\eta &=& \left[\begin{array}{c}
 \eta^0 \\
\eta^- \end{array}\right]=\left[\begin{array}{c}
\fr{1}{\sqrt{2}}(G_{V_1}+i G_{V_2}) \\
G^-_{V} \end{array}\right].\eea 
Component field masses take the form, $m_{V_1} = m_{V_2}= m_{V^\pm} \equiv m_V = g w/2$ and $m_{G_{V_1}} = m_{G_{V_2}}=m_{G_V}=0$, where the gauge boson doublet has a degenerate mass proportional to the gauge coupling $g$ and the breaking scale $w=\sqrt{2}\langle \chi^0\rangle$, due to $SU(2)_L$ conservation, whereas the Goldstone boson doublet associated with the gauge field has vanished mass due to the Goldstone theorem. If $\mathcal{S}$ is large, we may require more Higgs fields than $\chi^0$ to break it to $SU(2)_L$. However, $\chi^0$ is just enough to break the four generators that are associated with $V_\mu$, while the scalar multiplet under $\mathcal{S}$ that contains $\chi^0$ manifestly provides the corresponding Goldstone bosons, i.e. $\eta$. Here, note that $\chi^0$ is a $SU(2)_L$ singlet, impossibly being the Goldstone boson of a vector doublet. 

The gauge boson doublet is further separated in mass due to $SU(2)_L$ breaking. This mass separation is proportional to weak scales, resulting from potential interactions of $V_\mu$ with normal Higgs doublets, $\phi=(\phi^0, \phi^-)$ and $\phi'=(\phi'^0, \phi'^-)$, such as\footnote{The vector mass splitting happens only due to an interaction with adjoint scalar representation of $\mathcal{S}$, transforming similarly to the extended gauge field as well as containing the two doublets $\phi,\phi'$, whose couplings with $V_\mu$ are supplied by the gauge symmetry.} \bea \mathcal{L}_{\mathcal{S}} &\supset& \fr{1}{4}g^2 \left[(\phi^\dagger \phi)(V^\dagger_\mu V^\mu) + (\phi'^\dagger \phi')(V^\dagger_\mu V^\mu)\right.\crn 
&&\left.+(\phi^\dagger V^\mu)(V^\dagger_\mu \phi)+(\phi'^\dagger V^\mu)(V^\dagger_\mu \phi')\right.\crn
&&\left.-2(V^\dagger_\mu \phi)(V^{\mu\dagger} \phi')-2(\phi^\dagger V_\mu)(\phi'^\dagger V^\mu)\right],\eea whose couplings are set by $\mathcal{S}$ conservation. Substituting relevant vacuum values, we obtain \bea m^2_{V_1} &=& \fr{g^2}{4}\left[w^2+ \fr 1 2 (u-u')^2\right],\\ 
m^2_{V_2} &=& \fr{g^2}{4}\left[w^2+ \fr 1 2 (u+u')^2\right],\\
m^2_{V^\pm}&=& \fr{g^2}{4}\left[w^2+ \fr 1 2 (u^2+u'^2)\right],\eea where $u=\sqrt{2}\langle \phi^0\rangle$ and $u'=\sqrt{2}\langle \phi'^0\rangle$ are the mentioned $SU(2)_L$-breaking weak scales, satisfying $u,u'\ll w$ for consistency with the standard model. It is noted that the standard model Higgs field ($\varphi$) is embedded in $\phi,\phi'$ and others if available. Further, let us remind the reader that when the theory is quantized in a generic $R_\xi$-gauge, each Goldstone boson obtains a mass parameter proportional to the mass of the corresponding gauge boson due to the gauge fixing condition. 

Furthermore, assuming the $\mathcal{S}$ breakdown also contains two kinds of fermion singlets, 
\be N_{L},\hs N_{R},\ee which couple to lepton doublets, $l_{L}=(\nu_{L}, e_{L})$, through $V_\mu,\eta$, as well as they are coupled via $\chi^0$, such as
\bea 
\mathcal{L}_{\mathcal{S}}&\supset&-\fr{g}{\sqrt{2}}\bar{l}_{aL}\ga^\mu V_\mu N_{aL} + h_{ab}(\bar{l}_{aL} \eta + \bar{N}_{aL} \chi^0) N_{bR}\crn
&& -\fr 1 2 M_{ab} N_{a R} N_{b R}+H.c., \label{lbe1} \eea
where $a,b=1,2,3$ are family indices. Other couplings and masses if arisen due to $N_L$-$N_R$ interchange are not allowed by the $\mathcal{S}$ symmetry, for which $N_L$ and $l_{L}$, as well as $\chi^0$ and $\eta$, lie in the same multiplet of $\mathcal{S}$, whereas $N_R$ is a $\mathcal{S}$ singlet, as specified by the relevant terms in Eq.~(\ref{lbe1}). Additionally, being preserved by the above Lagrangian, the full gauge symmetry would provide a residual $Z_2$ parity, such that the new fields $V_\mu,\eta, N_{L,R}$ are odd, while the standard model fields and $\chi^0$ are even (cf. the next section). Without loss of generality, we assume $M_{ab}=\delta_{ak}\delta_{bk}M_k$ to be flavor diagonal, hence labeling the corresponding states to be $N_{kR}$, for $k=1,2,3$. We require $M_k\gg w$ so that $N_{kR}$ are manifestly decoupled from the $N_L$-$N_R$ mixing, i.e. $N_{kR}$ behave as physical eigenstates by themselves.        

\begin{figure}[h]
\includegraphics[scale=0.7]{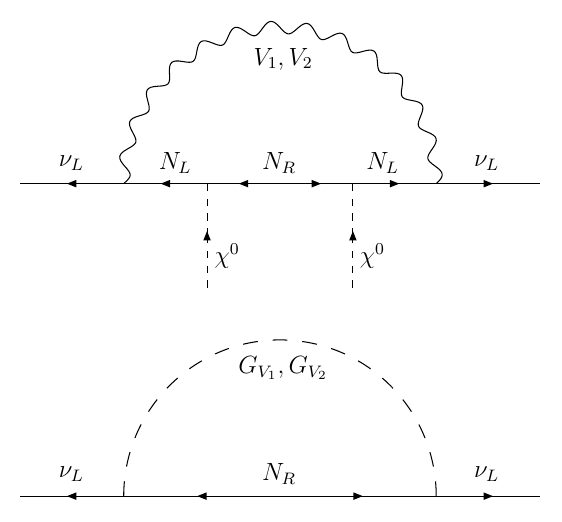}
\caption[]{\label{fig1} Scotogenic gauge mechanism for neutrino mass generation in a generic $R_\xi$-gauge, whereas there exists only the top diagram responsible for neutrino mass in the unitarity gauge according to $\xi\rightarrow \infty$.} 
\end{figure}
The Feynman diagrams that radiatively generate neutrino masses are depicted in Fig.~\ref{fig1}. Notice that the contribution of the Goldstone bosons in the bottom diagram is analogous to the inert scalars in the usual scotogenic setup \cite{Ma:2006km,Tao:1996vb}. However, since in this work they behave as the longitudinal components of the gauge fields, the neutrino masses are, by contrast, induced by the corresponding gauge vectors through the top diagram instead. The neutrino mass matrix that is derived in form of $\mathcal{L}_{\mathcal{S}}\supset -\fr 1 2 \bar{\nu}_{aL} (M_\nu)_{ab}\nu^c_{bL}+H.c.$ reads 
\bde\bea -i (M_\nu)_{ab}P_R&=&\int\fr{d^4 p }{(2\pi)^4}\left(-i\fr{g}{2}\ga^\mu P_L\right)\fr{i}{\slash\!\!\! p} \left(ih_{ak}\fr{w}{\sqrt{2}}P_R\right)\fr{i}{\slash\!\!\! p-M_k}\left(ih_{bk}\fr{w}{\sqrt{2}}P_R\right)\crn
&&\times \fr{i}{-\slash\!\!\! p}
\left(i\fr{g}{2}\ga^\nu P_R\right)\fr{-i}{p^2-m^2_{V_1}}\left(g_{\mu\nu}-\fr{(1-\xi)p_\mu p_\nu}{p^2-\xi m^2_{V_1}}\right)\crn
&&+\int\fr{d^4 p }{(2\pi)^4}\left(\fr{g}{2}\ga^\mu P_L\right) \fr{i}{\slash\!\!\! p} \left(ih_{ak}\fr{w}{\sqrt{2}}P_R\right)\fr{i}{\slash\!\!\! p-M_k}\left(ih_{bk}\fr{w}{\sqrt{2}}P_R\right)\crn
&&\times \fr{i}{-\slash\!\!\! p}
\left(\fr{-g}{2}\ga^\nu P_R\right)\fr{-i}{p^2-m^2_{V_2}}\left(g_{\mu\nu}-\fr{(1-\xi)p_\mu p_\nu}{p^2-\xi m^2_{V_2}}\right)\crn
&&+\int\fr{d^4 p }{(2\pi)^4}\left(i\fr{h_{ak}}{\sqrt{2}} P_R\right) \fr{i}{\slash\!\!\! p-M_k}
\left(i\fr{h_{bk}}{\sqrt{2}}P_R\right)\fr{i}{p^2-\xi m^2_{V_1}}\crn
&&+\int\fr{d^4 p }{(2\pi)^4}\left(-\fr{h_{ak}}{\sqrt{2}} P_R\right) \fr{i}{\slash\!\!\! p-M_k}
\left(-\fr{h_{bk}}{\sqrt{2}} P_R\right)\fr{i}{p^2-\xi m^2_{V_2}}\crn
&=& -\fr{3P_R}{2}h_{ak}h_{bk} M_k\int \fr{d^4p}{(2\pi)^4}\fr{m^2_{V_1}-m^2_{V_2}}{(p^2-M^2_k)(p^2-m^2_{V_1})(p^2-m^2_{V_2})}.\eea \ede The contributions of the real ($V_1$) and imaginary ($V_2$) parts of $V^0$ have opposite signs, because the product of the two neutrino vertices according to each vector field changes the sign. This is also valid for the Goldstone boson contributions. The result that is a difference of the two kinds of contributions depends on the $V_{1,2}$ mass splitting. It is clear that the sum of the $V_1$ and $G_{V_1}$ contributions, as well as that of $V_2$ and $G_{V_2}$, is independent of the gauge fixing parameter $\xi$, coinciding with the top diagram contribution in the unitarity gauge.

Given the parameters that obey $x\gg y,z$, we have \bea && \int \fr{d^4 p}{(2\pi)^4}\fr{1}{(p^2-x )(p^2-y)(p^2-z)}\crn
&&\simeq \fr{i}{16\pi^2}\fr{y\ln y/x-z \ln z/x}{x (y-z)}.\eea Applying to the neutrino mass matrix due to $M_k\gg m_{V_1},m_{V_2}$ yields a final form,
\bea (M_\nu)_{ab} &\simeq& \fr{3}{32\pi^2}\fr{h_{ak}h_{bk}}{M_k}\left(m^2_{V_1}\ln \fr{m^2_{V_1}}{M^2_k}-m^2_{V_2}\ln \fr{m^2_{V_2}}{M^2_k}\right)\crn
&\simeq& -\fr{3g^2}{64\pi^2} \fr{h_{ak}h_{bk} u^2}{M_k}\left(\ln \fr{m^2_V}{M^2_k}+1\right),\eea where the last approximation uses the small mass splitting $m^2_{V_2}-m^2_{V_1}= g^2 u^2/2\ll m^2_V\equiv g^2 w^2/4$. Here, the adjoint representation that contains $\phi,\phi'$ has been assumed to be Hermitian similar to the $\mathcal{S}$ gauge field, which follows that $\phi'=\phi$, thus $u'=u$, for simplicity. The seesaw scale is significantly reduced by $3g^2/64\pi^2$, in agreement to Ref. \cite{Ma:2006km,Tao:1996vb}. Assuming $u\sim 100$ GeV, $g\sim 0.65$, and $h\sim 10^{-3}$, the neutrino mass is recovered to be $M_\nu\sim 0.1$ eV, if $M_k\sim 500$ TeV. This seesaw scale coincides with the condition $M_k\gg w$, where $w\sim 10$ TeV results from the collider constraint for a $\mathcal{S}$ breaking. 

The dark matter candidate is the lightest of $V_{1,2}$ and $N_{1,2,3L}$, stabilized by the residual gauge symmetry, $Z_2$. Notice that $N_{1,2,3L}$ mass is given by a seesaw proportional to $\sim (hw)^2/M_k$, which is radically smaller than ${V_{1,2}}$ mass. Thus, the dark matter candidate is a sterile neutrino, $N_{L}$, composed of $N_{1,2,3L}$.  

An alternative case that is not discussed throughout this work is that we only necessarily impose $M_k\gg h w$, not $M_k\gg w$, since $h$ is small. In such case, one can take $M_k \sim w$, hence $N_{k R}$ and $V_{1,2}$ equivalently contributing to the neutrino mass. Then, the neutrino mass gets the correct value if chosen $M_k\sim m_{V}\sim 10$ TeV for $h\sim 10^{-4}$. The dark matter phenomenology is analogous to the previous case, so it should be skipped. 

Last, but not least, if $M\sim h w$, in contrast with the previous cases, the $N_L$-$N_R$ mixing is finite. One must diagonalize a $6\times 6$ mass matrix of $(N_L,N_R)$ to identify physical sterile-fermion eigenstates $N'_j$ with respective masses $M'_j$, for $j=1,2,3,\cdots,6$, which are related to the original states as $(N_L,N^c_R)=(U,W) N'$. In this case, the results obtained above are straightforwardly generalized. The six fields $N'_{1,2,3,\cdots,6}$ contribute to neutrino mass, in addition to $V_{1,2}$, such as  
\bea (M_\nu)_{ab} &=& \fr{3(hW)_{aj} (hW)_{bj} M'_j}{32\pi^2}\crn
&&\times \left(\fr{m^2_{V_1}\ln \fr{m^2_{V_1}}{M'^2_j}}{M'^2_j-m^2_{V_1}}-\fr{m^2_{V_2}\ln \fr{m^2_{V_2}}{M'^2_j}}{M'^2_j-m^2_{V_2}}\right).\eea The dark matter candidate is the lightest of $V_{1,2}$ and $N'_{1,2,3,\cdots,6}$, which all have a mass at TeV dependent on the magnitude of $W$ mixing elements. That said, the model implies a vector dark matter with a mass at TeV scale in the generic case. Once a gauge completion is given, the vector dark matter observables are predicted, similar to the 3-$P$-1-1 model \cite{Nam:2020twn}.      

\section{\label{sgc} Discussion of gauge completion and matter parity} 

The $V,\eta$ doublets are associated with the quotient group, $\mathcal{S}/SU(2)_L$, determined when $\mathcal{S}$ is broken down to $SU(2)_L$. The most minimal extension of $SU(2)_L$ is \be \mathcal{S}=SU(3)_L,\ee a higher weak isospin symmetry, which has the adjoint representation $A_\mu\sim \underline{8}$ decomposed into $SU(2)_L$ as \be \underline{8}=\underline{3}\oplus \underline{2} \oplus \underline{2}^* \oplus \underline{1}.\label{21}\ee The adjoint representation possesses components,
\bea A_\mu &=& T_n A_{n \mu}\crn
&=& \fr 1 2 \left[\begin{array}{ccc} 
A_{3\mu} +\fr{1}{\sqrt{3}}A_{8\mu} & \sqrt{2} W^+_\mu & \sqrt{2} V^0_\mu \\
\sqrt{2} W^-_\mu & -A_{3\mu} +\fr{1}{\sqrt{3}}A_{8\mu} & \sqrt{2} V^-_\mu \\
\sqrt{2} V^{0*}_\mu & \sqrt{2} V^+_\mu & \fr{-2}{\sqrt{3}}A_{8\mu} \\
\end{array}\right].\eea The $SU(2)_L$ multiplets, $\underline{3}$ and $\underline{1}$, include $W^\pm, A_{3,8}$ determined by the $T_{1,2,3,8}$ generators of $SU(3)_L$, while the $SU(2)_L$ doublets, $\underline{2}$ and $\underline{2}^*$, are just $V$ defined by the $T_{4,5,6,7}$ generators of $SU(3)_L$. 

The $\mathcal{S}$ breakdown to $SU(2)_L$ is proceeded by a scalar triplet, $\Phi\sim \underline{3}=\underline{2}\oplus \underline{1}$ under $SU(2)_L$, where $\underline{2}$ is the Goldstone boson doublet, $\eta$, while $\underline{1}$ is the new Higgs field, $\chi$, that has the VEV, $w$, namely 
\be \Phi = \left[
\begin{array}{c}
\fr{1}{\sqrt{2}}(G_{V_1}+i G_{V_2}) \\
G^-_{V}\\
\fr{1}{\sqrt{2}}(w+\mathcal{H}+i G_{A_8})
\end{array}\right].\label{adnt1}\ee Hence, $m_V=gw/2$ is set by the VEV of $\Phi_3$, where $g$ is the $SU(3)_L$ coupling constant to be identical to the $SU(2)_L$ one below the breaking scale. Additionally, this VEV breaks the $T_{4,5,6,7,8}$ generators as suitable, in which $G_{A_8}$ is eaten by the gauge field $A_{8\mu}$, similar to the Goldstone boson doublet as eaten by $V_\mu$. Notice that the vacuum structure in (\ref{adnt1}) is derived by a potential $V(\Phi)=\mu^2_\Phi \Phi^\dagger \Phi +\fr 1 2  \la_\Phi (\Phi^\dagger \Phi)^2$ with $\mu^2_\Phi<0$ and $\la_\Phi>0$ for which $w=\sqrt{-2\mu^2_\Phi/\la_\Phi}$, where $\mu_\Phi$ and $\la_\Phi$ match those for $\eta$, i.e. $\mu_\eta$ and $\la_\eta$, at low energy, respectively.  

Under the $\mathcal{S}$ symmetry, the lepton doublets $l_{aL}=(
\nu_{aL}, e_{aL})$ must be enlarged to become its appropriate representations. The simplest of which is they are put in triplet, $\underline{3}=\underline{2}\oplus \underline{1}$, as 
\be \psi_{aL}=\left[\begin{array}{c}
\nu_{aL}\\
e_{aL}\\
N_{aL}\end{array} \right],\ee where $N_{aL}$ are $SU(2)_L$ singlet, $\underline{1}$. Further, assume that $N_{aL}$ have right-handed partners, $N_{aR}$, and all possess $Q=0=B-L$. All the right-handed leptons, $N_{aR}$ and $e_{aR}$, are retained as $SU(3)_L$ singlet.  

The Lagrangian invariant under $\mathcal{S}$ includes  
\bea \mathcal{L}_{\mathcal{S}} &=& \bar{\psi}_{aL} i \ga^\mu D_\mu \psi_{aL} +(D^\mu \Phi)^\dagger (D_\mu \Phi)  -\fr 1 2 \mathrm{Tr} A^\mu A_\mu \crn
&& -V(\Phi) +[ h_{ab} \bar{\psi}_{aL}\Phi N_{bR} -\fr 1 2 M_{ab} N_{a R} N_{b R}+H.c.],\nn\eea
where $D_\mu =\pa_\mu + ig A_\mu$ is the covariant derivative and $e_{aR}$ do not have couplings as $N_{aR}$ due to the conservation of electric charge. It further contains 
\bea 
\mathcal{L}_{\mathcal{S}}&\supset&-\fr{g}{\sqrt{2}}\bar{l}_{aL}\ga^\mu N_{aL} V_\mu + h_{ab}\bar{l}_{aL} N_{bR} \eta \crn
&&+ h_{ab}\fr{w}{\sqrt{2}} \bar{N}_{aL} N_{bR}-\fr 1 2 M_{ab} N_{a R} N_{b R}+H.c., \eea where $h_{ab}$ would be fixed by the Goldstone boson equivalence theorem. This matches the above Lagrangian which governs the scotogenic gauge mechanism.    

The mentioned gauge symmetry would provide a matter parity, $Z_2$, such that all $V,\eta,N_L$ are odd, while every standard model field is even, conserved by the above Lagrangian \cite{Dong:2013wca}. Last, but not least, it is easily verify that a scalar octet that contains $\phi,\phi'$ similar to (\ref{21}) for which $\phi\sim \underline{2}$ and $\phi'^*\sim \underline{2}^*$ would separate the masses of $V_{1,2}$. Of course this scalar contributes to the total Lagrangian in addition to that of $\Phi$. 

\section{\label{dmlfv} Remark on dark matter observables and charged lepton flavor violation processes} 

The Yukawa coupling $h$ is somewhat smaller than the gauge coupling $g$, as implied by the neutrino mass generation. Hence, the gauge portal would dominantly contribute to dark matter observables and charged lepton flavor violation processes, which are completely determined, given the above gauge completion.

Concerning dark matter, we assume $V_1$ to be lightest among the dark fields. Hence, it is stabilized by the matter parity to be a dark matter candidate. In the early universe, the dark matter candidate annihilates to electroweak bosons and usual fermions, such as \be V_1V_1\to W^+W^-, ZZ, HH, f\bar{f},\ee which are all set by the gauge coupling. Notice that dark matter annihilations to $W^+W^-$ and $ZZ$ obey the unitarity condition \cite{Nam:2020twn}. Hence, the total dark matter annihilation cross-section is manifestly suppressed by $\langle \sigma v_{\mathrm{rel}}\rangle_{V_1}\sim 1/m^2_{V_1}$. The dark matter relic density obtains a correct value, $\Om_{V_1} h^2\simeq 0.1\ \mathrm{pb}/\langle \sigma v_{\mathrm{rel}}\rangle_{V_1}\simeq 0.12$ \cite{ParticleDataGroup:2022pth}, if $V_1$ gets a mass in TeV regime. Since $V_{1,2}$ are separated in mass, the dark matter $V_1$ avoids a dangerous scattering with nuclei via $Z$ exchange in direct detection \cite{Barbieri:2006dq}. Conversely, this scattering proceeds via $H$-exchange, which induces an effective coupling $\mathcal{L}_{\mathrm{eff}}\supset (g^2 m_q/m^2_H)V^\mu_1V_{1\mu} q\bar{q}$. This leads to a spin-independent (SI) cross-section of $V_1$ with nucleon ($N=p,n$), such as
\bea \sigma^{\mathrm{SI}}_{V_1-N} &=& \fr{g^4m_N^4 f^2_N}{4\pi m^4_H m^2_{V_1}}\crn
&\simeq&3\times 10^{-46}\times \left(\fr{3\ \mathrm{TeV}}{m_{V_1}}\right)^2\ \mathrm{cm}^2,\eea  where $f_N$ parametrizes Higgs-nucleon coupling, taking the value $f_N\simeq 0.35$ \cite{Ellis:2000ds}, and we have used $m_N\simeq 1$~GeV, $m_H\simeq 125$ GeV, and $g=0.65$. This prediction of the SI cross-section agrees with the latest measurement \cite{LZ:2022lsv} for dark matter mass at $m_{V_1}\sim 3$ TeV, as expected.

\begin{figure}[h]
\bc
\includegraphics[scale=1]{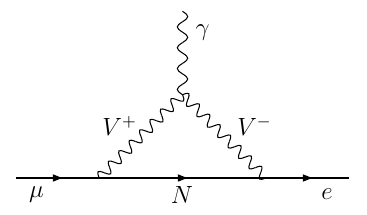}
\caption[]{\label{fig2} Dark field contribution to $\mu\to e\gamma$.}
\ec
\end{figure}

In the standard model, lepton flavor is conserved as far as neutrinos are massless. The observed neutrino oscillations are a direct evidence that lepton flavor is violated. This flavor violation in neutrino propagator suggests that it would exist in charged lepton flavor, e.g. $\mu\to e\gamma$. The observation of charged-lepton flavor violation would be important indication for new physics, enhancing our understanding of lepton sector. Concerning charged-lepton flavor violation in this model, the most dangerous process is associated with $\mu\to e \gamma$ through exchange of dark fields, as depicted in Fig. \ref{fig2}. It is easily to derive \cite{Lindner:2016bgg} 
\be \mathrm{Br}(\mu\to e \gamma)\simeq  2.2\times 10^{-10}\times \left(\fr{3\ \mathrm{TeV}}{m_{V^\pm}}\right)^4\times |U_{\mu N}U^*_{eN}|^2,\ee where $U_{lN}=(V^\dagger_{L} V'_{L})_{lN}$   is $lN$ element of left-handed mixing matrix product of charged lepton $V_L$ and sterile neutrino $V'_L$. The current sensitivity reported by the MEG experiment implies $\mathrm{Br}(\mu \to e\ga)\simeq 4.2\times 10^{-13}$ \cite{MEG:2016leq}. This translates to $|U_{\mu N}U^*_{eN}|\sim 10^{-2}$ for $m_{V^\pm}\sim 3$ TeV.

\section{\label{concl} Conclusion}         

We have given a review of the scotogenic mechanism. We have indicated that this mechanism can be gauged, yielding a novel scotogenic gauge scheme that explains neutrino mass and dark matter. An alternative prediction is that the lightest of dark vector, i.e. $V_{1}$, may be a dark matter candidate, which can be detected in current high energy colliders or at direct dark matter detection. The dark fields also contribute to $\mu \to e \gamma$ with a rate consistent with the current bound.   

\section*{Acknowledgements}

The work is funded by Vietnam National Foundation for Science and Technology Development (NAFOSTED) under grant number 103.01-2023.50.

\bibliographystyle{apsrev4-1}
\bibliography{combine}

\end{document}